\documentclass[ twocolumn,aps,prd,
               preprintnumbers,numbers,sort&compress,
               nofootinbib,
                            showpacs,
               colorlinks,
               linkcolor=blue,
               citecolor=blue,
               superscriptaddress]{revtex4-1}
   \newcommand{\exclude}[1]{}

\usepackage{epsfig,graphicx}
\usepackage{amsmath,amssymb,bm}
\usepackage{psfrag}
\usepackage{feynmp}
\usepackage{hyperref}

\newcommand{\beq}{\begin{equation}}
\newcommand{\eeq}{\end{equation}}
\newcommand{\be}{\begin{eqnarray}}
\newcommand{\ee}{\end{eqnarray}}

\newcommand{\bb}{\bibitem}

\def\d{\partial}
\def\+{\dagger}
\def\la{\langle}
\def\ra{\rangle}
\def\<{\langle}
\def\>{\rangle}

\newcommand{\Lqcd}{\Lambda_{\mathrm{QCD}}}

\begin{document}
\title{Cosmological Magnetic Field and  Dark Energy as two sides of the same coin.}

\author{Ariel R. Zhitnitsky}

\affiliation{Department of Physics and  Astronomy, University of British Columbia, Vancouver, B.C. V6T 1Z1, Canada}

\begin{abstract}
 It has been recently argued \cite{Barvinsky:2017lfl} that the de Sitter phase in cosmology might be naturally  generated as a result of  dynamics of the  topologically  nontrivial  sectors in a   strongly coupled QCD-like  gauge theory  in expanding universe. It is known that the de Sitter phase is realized  in the history of our Universe twice: the first occurrence  is coined as the inflation,  while  the  second time (which is occurring  now) is dubbed  as the  dark energy (DE). The crucial element of the  proposal \cite{Barvinsky:2017lfl} is the presence     of a nontrivial  
   gauge holonomy  which is the source of the vacuum energy     leading    to the de Sitter behaviour.  It has been also argued  that the anomalous coupling of the system with   the Standard Model (SM)   particles   leads to the reheating epoch in case  of the {\it inflationary} phase.  
   A similar anomalous coupling of the system with the Maxwell $E\&M$ field  during the {\it DE epoch}  generates   the cosmological    magnetic field. The   intensity of the field is estimated on the level of   $10^{-10}$G while  the corresponding correlation length reaches the scale of   the  visible Universe.

  \end{abstract}
 
\maketitle

\section{Introduction}\label{introduction}
This work is mostly motivated by recent proposal \cite{Barvinsky:2017lfl} where the vacuum energy (and accompanying it the de Sitter phase) is dynamically generated. The  proposal  \cite{Barvinsky:2017lfl} can be viewed as a synthesis of two naively unrelated   ideas  discussed previously in refs. \cite{Barvinsky:2006uh,Barvinsky:2006tu,why} and  \cite{Zhitnitsky:2013pna,Zhitnitsky:2014aja,Zhitnitsky:2015dia} correspondingly.    On the gravity side \cite{Barvinsky:2006uh,Barvinsky:2006tu,why} the nontrivial element of the construction is represented by the Euclidean spacetime with a time compactified to a circle $\mathbb{S}^1$. On the gauge field theory side  \cite{Zhitnitsky:2013pna,Zhitnitsky:2014aja,Zhitnitsky:2015dia} the same  $\mathbb{S}^1$ plays a crucial role when the gauge configurations may assume a nontrivial holonomy along  $\mathbb{S}^1$. Precisely the gauge configurations with the nontrivial  holonomy
along $\mathbb{S}^1$ may  serve as a source of vacuum energy density, which eventually leads to the de Sitter behaviour.   

The focus of proposal \cite{Barvinsky:2017lfl}  coined as the  ``holonomy inflation" was the study of the vacuum energy and the corresponding  de Sitter behaviour  in application to the inflationary Universe.  It has been also  suggested  in that proposal that the holonomy inflation ends as a result of anomalous coupling of the   system with massless Standard Model (SM) gauge fields with known coefficients.  

The present work applies the same ideas on dynamical generation of the vacuum energy to the dark energy (DE) epoch when the corresponding strongly coupled gauge theory is well known, it is QCD   characterized by a single dimensional parameter, $\Lqcd\sim 0.1$ GeV. A similar anomalous coupling (which was the source of the reheating in the ``holonomy inflation"  in  \cite{Barvinsky:2017lfl} when the vacuum energy is transferred to the massless gauge fields)    generates  the cosmological    magnetic helical configurations with enormous correlation length reaching the size of the entire visible Universe during the present    DE  epoch.    The focus of the present work is an analysis  of the   generation  mechanism      of such  long ranged magnetic  field. 

Before we proceed with outline of this work we would like to make few remarks   
on conventional approaches to study the cosmological magnetic field. We refer to the classical  review papers \cite{kronberg,grasso} and more recent review \cite{durrer} for details and references.  It is normally assumed that
magnetic fields in astronomical structures of different sizes, from stars  $R\sim10^{11} {\rm cm}$ to galaxy clusters $R\sim10^{24} {\rm cm}$ are produced by amplification of pre-existing   weaker ``seed"  magnetic fields via different types of dynamo. Two broad classes of models for the origin of the seed fields are discussed:  1. primordial magnetic field (seeds)  is produced during different dramatic events in evolution of the Universe such as inflation, electroweak  phase transition, QCD transition, i.e.  during the epochs preceding the structure formation; 2. the process of generation of the seed magnetic fields accompanies the gravitational collapse leading to structure formation. We shall not comment on many problems related to this conventional picture referring to the reviews   \cite{kronberg,grasso,durrer}. 

The unorthodox mechanism we are advocating in the present work is drastically distinct from previous conventional approaches.  
Essentially, the magnetic field in our framework is generated with enormous  scale  from the  moment when it was born  as the source of  its energy is  the  DE occupying the entire Universe. Therefore,  there is no need for amplification nor for different types of inverse cascades as the correlation length of the produced field is already  characterized by the largest possible scale. 
The intensity of this correlated magnetic field  is estimated on the level of $ B\sim 10^{-10}$ G, and the intensity of the field  $B^2$ is  proportional 
to the DE density  $ \rho_{\rm DE} \approx (2.3\cdot 10^{-3} \text{eV})^4$ with  calculable (in principle) coefficient.

This intensity is very close to the upper limit, but not ruled out. In fact such fields can be studied by future UHECR telescopes, see Fig. 14 in ref.  \cite{durrer}. 

Our presentation is organized  as follows. In next section \ref{topology} we overview the basic ideas and results on the nature of vacuum energy from ref. \cite{Barvinsky:2017lfl}.  The nature of the DE plays a crucial  role in our   framework as it is the source of the cosmic magnetic field, which is the main subject of the present work. Therefore, we overview the basic ideas of  \cite{Barvinsky:2017lfl} in context of the present work in great details for benefits of the readers.
 In Sect.\ref{anomalous-coupling} we explain how the DE couples to the EM   field through the chiral anomaly. Precisely this coupling is responsible for the generation of the long ranged magnetic  field, which is the subject of Section \ref{EM-generation}
 where we estimate its intensity.  We conclude in Section\ref{conclusion} with few comments on future development, and possible observational tests  which may support or rule out this new paradigm when DE and cosmic magnetic field represent two sides of the same coin and are produced  at the same epoch. 

\section{The topology as the source of the gravitating vacuum energy}\label{topology}
The goal here is to overview the basic ideas advocated in  \cite{Barvinsky:2017lfl}, see also a number of precursor  references therein.   

  In approach   \cite{Barvinsky:2017lfl} the vacuum energy entering the Friedmann equation  is defined as  $\Delta\rho\equiv \rho_{\rm FRW} -\rho_{\mathrm{Mink}}$. This definition  for the vacuum energy for the first time was advocated   in 1967   by Zeldovich~\cite{Zeldovich:1967gd} who argued that  $\rho_{\text{vac}}=\Delta\rho \sim Gm_p^6 $ must be proportional to the gravitational constant with $m_p$ being the proton's mass. Later on such  definition for the relevant energy $\Delta\rho\equiv \rho_{\rm FRW} -\rho_{\mathrm{flat}}$ which  enters the Einstein equations has been advocated from   different perspectives in a number of papers written by the researches from different fields, including particle physics, cosmology, condensed matter physics.
This subtraction prescription is consistent with conventional  subtraction procedure of the divergent ultra local bare cosmological constant because in the infinitely large  flat space-time the corresponding contribution is proportional to the  $\delta^4(x)$ function as explained in  \cite{Barvinsky:2017lfl}. At the same time  the  nontrivial correction to    $\Delta\rho$ as discussed below is a   non-local function of the geometry and cannot be renormalized by any UV counter-terms.

  In the present work  we consider  the geometry  $\mathbb{R}^3\times \mathbb{S}^1$
  instead of FRW geometry to simplify the arguments. The key element in this framework is the presence of a dimensional parameter ${\cal{T}}^{-1}$ which 
 plays the role of the Hubble constant $H$ in FRW geometry which distinguishes  FRW geometry from flat infinite space-time geometry. In other words,
 we have a dimensional parameter  $\cal{T}$ which is assumed to be order  $\sim H^{-1}$ 
 and which parametrizes the difference between   nontrivial   and  trivial (flat infinite space-time $ \mathbb{R}^4 $) geometries. In computations  \cite{Barvinsky:2017lfl} parameter $\cal{T}$ is the proper length of the $\mathbb{S}^1$-period. 
 %Our prescription  $\Delta\rho\equiv [\rho_{\rm FRW} -\rho_{\mathrm{flat}}]$   implies that  the vacuum energy which enters  the  Friedmann  equation   is  $\rho\equiv[\rho({\cal{T}}^{-1})-\rho(0)]$, where $\rho({\cal{T}}^{-1})$ is the energy of the gauge field holonomy on a compactified spacetime coordinate of length $\cal{T}$.   
 As we already mentioned, this prescription (when $\Delta\rho\equiv [\rho_{\rm FRW} -\rho_{\mathrm{flat}}]$ is identified with physical energy, similar to the Casimir Effect)  is consistent with the  Einstein equations when the vacuum energy  approaches zero,  $\Delta\rho\rightarrow 0$  for the flat  space-time   which itself may be considered as a limiting case with ${\cal{T}}\rightarrow\infty$.

The key element of the framework \cite{Barvinsky:2017lfl}  is that  the vacuum  energy 
receives the linear correction ${\cal{T}}^{-1}$ at large $\cal{T}$ in contrast with naively expected quadratic corrections ${\cal{T}}^{-2}$ such that the vacuum energy entering the Friedmann equation assumes the form 
\be
\label{delta}
&& \rho_{\rm DE}\equiv \Big(E_{\rm vac}[ \mathbb{R}^3 \times \mathbb{S}^1]-E_{\rm vac}[ \mathbb{R}^3 \times \mathbb{R}^1]\Big)= \Lqcd^3 \frac{\bar{c}_{{\cal{T}}}}{{\cal{T}}}, ~~~
\ee
where the vacuum energy can be represented as follows 
\be
\label{E_vac}
&&E_{\rm vac}[ \mathbb{R}^3 \times \mathbb{S}^1]\simeq   -\frac{32\pi^2}{g^4} \Lqcd^4  \left(1-   \frac{c_{{\cal{T}}}}{{\cal{T}} \Lqcd} \right)\nonumber\\
&&\simeq -\frac{32\pi^2}{g^4}  \Lqcd^4 +   \Lqcd^3 \frac{\bar{c}_{{\cal{T}}}}{{\cal{T}}}+{\cal{O}}(\frac{1}{{\cal{T}}^2}). 
\ee
In this expression  we redefined $\bar{c}_{{\cal{T}}}\equiv\frac{32\pi^2}{g^4}c_{{\cal{T}}}$. The   coefficient  $c_{{\cal{T}}}\sim 1$ is, in principle, a calculable parameter\footnote{It can be in principle computed in strongly coupled QCD using the lattice Monte Carlo simulations, similar to studies \cite{Yamamoto:2014vda}.},    expected to be order of one. The linear dependence ${\cal{T}}^{-1}$ of the relevant portion of the vacuum energy (\ref{delta})   on external parameter ${\cal{T}}^{-1}\sim H$ suggests   that  $ \rho_{\rm DE}$ numerically is very close to the observed value today, i.e.
\be
\label{delta1}
 \rho_{\rm DE} \simeq   \Lqcd^3 \frac{\bar{c}_{{\cal{T}}}}{{\cal{T}}}\sim  \Lqcd^3 H \sim \left(10^{-3} {\rm eV} \right)^4.
\ee
One  should   also mention that  this numerical coincidence in estimate  (\ref{delta1}) was the main  motivation    to advocate the proposal \cite{Urban:2009vy,Urban:2009yg} that the  driving force for the dark energy  is  a nontrivial  dynamics of the    topological sectors in strongly coupled QCD (admittedly, without much deep   understanding behind the formula at that time).  

Few important comments regarding formulae (\ref{delta}) and (\ref{E_vac}) are in order. 

1.   All computations  leading to (\ref{E_vac}) are performed in the Euclidean space-time where the relevant gauge configurations describing the tunnelling processes are defined. Using this technique one can  compute the energy  density $\rho$ and the pressure $P$ in the Euclidean space. As usual, we assume that there is analytical continuation to Lorentizan space-time where the physical energy density has  the same form.
  In our context it means that the parameters $P, \rho$ and equation of state (EoS) as given by (\ref{EoS}) below are interpreted as the corresponding parameters in physical Lorentizan  space-time. 
    
  2. The same arguments also suggest that the parameter ${\cal{T}}$ entering (\ref{E_vac}) is 
  a constant parameter of the system (not to be confused with observed Hubble $H_{\rm obs} (t)$ which is time dependent in FRW Universe).  The cosmological evolution in the Lorentizan  space-time  is determined by  the analytic continuation  as discussed  in ref.  \cite{Barvinsky:2017lfl}. 

3. What is the interpretation of the parameter  ${\cal{T}}$ in physical Lorentizan  space-time?
In the system with Euclidean signature  the parameter  ${\cal{T}}$ is determined by the size of $\mathbb{S}^1$, which is normally can be interpreted as the inverse temperature of the system in Lorentizan  space-time. We think it is a proper interpretation even though there is no any thermodynamical processes which are occurring and characterized  by extremely low  temperature ${\cal{T}}^{-1}\sim H\sim 10^{-33} {\rm eV}$.

4. The vacuum energy $E_{\rm vac}$ is defined in conventional way in terms of the path integral. It has a 
 ``non-dispersive"    nature, which implies that the corresponding vacuum energy cannot be expressed in terms of conventional propagating degrees of freedom (absorptive part) using the dispersion relations to compute the dispersive part.  
  Furthermore,  all effects represented by eq. (\ref{E_vac}) are obviously non-analytical  in coupling constant $\sim \exp(-1/g^2)$ and can not be seen in perturbation theory\footnote{\label{ghost}This non-dispersive nature of the vacuum energy is well known 
to the QCD community: it appears in computation of the topological susceptibility (which is expressed as  the second derivative of the vacuum energy with respect to $\theta$).  The corresponding non-dispersive contact term was postulated by Witten in \cite{witten}, while the same term with a ``wrong sign" in the correlation function was  saturated by the Veneziano ghost in \cite{ven,vendiv}, see Appendix A1 in  \cite{Barvinsky:2017lfl} for references and  details.}.
  %These arguments obviously suggest that there is no any local effective field $\Phi(x)$ (inflaton) which could describe  these features of the vacuum energy in gauge theories. These arguments are obviously consistent with our  discussions in previous  subsection \ref{holonomy}. 
  Non-analytical structure emerging in eq. (\ref{E_vac}) can be easily understood without precise computations. Indeed,   $ \Lqcd$ in this formula appears as a result of tunnelling events which always proportional to $\Lqcd^4\sim \exp(-S_{\rm cl})\sim \exp(-1/g^2)$, while 4 zero modes which   accompany  every magnetic  monopole constituent  (see item 5 below)   
 of the classical caloron solution with nontrivial holonomy  produces the factor $\sim [\sqrt{S_{\rm cl}} ]^4\sim g^{-4}$, see \cite{Zhitnitsky:2015dia} for the details and references. 

5. One can view the relevant topological Euclidean configurations which saturate (\ref{E_vac})   as the 3d     magnetic monopoles  wrapping around  $\mathbb{S}^1$ direction. These configurations are characterized by the  non-vanishing holonomy, which eventually generates the linear  (rather than quadratic) correction $\sim1/{\cal{T}}$ to the vacuum energy density. For the specific geometry (leading to the de Sitter behaviour) considered in \cite{Barvinsky:2017lfl}   the parameter ${\cal{T}}$ and $H$ are related ${\cal{T}}\simeq \pi/H$ such that   $\Delta \rho\sim H$ when Hubble parameter is a  constant\footnote{\label{H}A nonzero holonomy for the vacuum configurations saturating the vacuum energy represents  a technical explanation   why the conventional argument (that the correction in (\ref{E_vac}) must be quadratic in $H^2$ in gravitational  background  rather than linear in $H$)  fails. The point is that the holonomy is independent    gauge invariant non-local characteristic of the system,  similar to the Polyakov's line, which cannot be expressed in terms of the local curvature $R$, which is indeed is quadratic in $H$ as $R\sim H^2$.   Explicit computations in Hyperbolic space support this claim, see item 8.}.

6. In the cosmological context such configurations are highly unusual objects: they obviously describe the non-local physics
  as the holonomy  is a nonlocal object. Indeed,  the holonomy  defines the dynamics  along the entire history  of evolution of the system.  This entire gauge configuration is a mere  saddle point in Euclidean   path integral computation which describes the instantaneous tunnelling event, rather than propagation of  a physical  degree of freedom.
  
 7. The generation of the ``non-dispersive" energy $E_{\rm vac}$  is highly non-local   effect as it is saturated by the gauge configurations with nontrivial holonomy.   Precisely this feature of non-locality implies that the relevant energy $\Delta \rho$ which enters the Friedmann equation (\ref{delta1}),   cannot be expressed in terms of a gradient expansion in any effective local field theory.
 \exclude{
 8. Our subtraction prescription as explained in section \ref{holonomy} is consistent will all fundamental principles  of QFT.
 What is more important is that the correction   to the energy $\Delta \rho$ which enters the Friedmann equation (\ref{Delta1}),  cannot be renormalized by any UV counter-terms as it is generated by non-local configurations.
}

  8. The basic  idea of the framework \cite{Barvinsky:2017lfl} on dynamical generation of the vacuum energy leading to the de Sitter behaviour is that there is a linear correction (with respect to the inverse size of the system) to the energy 
     \be
\label{ratio1}
\frac{E_{\rm vac}[ \mathbb{R}^3 \times \mathbb{S}^1]}{E_{\rm vac}[ \mathbb{R}^3 \times \mathbb{R}^1]}
 \simeq  \left(1-   \frac{c_{{\cal{T}}}}{{\cal{T}}\Lqcd} \right). 
\ee
  This  correction  $\sim{\cal{T}}^{-1}$ is generated  in spite of the fact that the system has  a gap $\Lqcd$ which naively implies that the system must not be sensitive to  the  size ${\cal{T}}$ of the system at all.  We already mentioned that the correction ${\cal{T}}^{-1}$ is nevertheless generated   because the vacuum energy (\ref{delta}), (\ref{delta1}) has a ``non-dispersive" nature,   not associated with any propagating massive degrees of freedom, but rather is related to instantaneous tunnelling events (expressed in terms of the Veneziano ghost, mentioned in footnote \ref{ghost}, as the presence of the topologically protected pole). Explicit computations in hyperbolic space $\mathbb{S}^1\times\mathbb{H}^3$
   \cite{Zhitnitsky:2015dia}
  and simplified ``deformed QCD" model \cite{Thomas:2012ib}, along with the lattice simulations  \cite{Yamamoto:2014vda} support this claim. 
  
  What is an intuitive way to understand the effect? Imagine that we study the Aharonov-Casher effect.
We insert an external charge into a superconductor when  the electric field $E$ is screened, i.e. $E\sim Q\exp(-r/\lambda)$ with $\lambda $ being the penetration depth.
Nevertheless, a neutral magnetic fluxon will be still sensitive to an inserted external charge $Q$ at arbitrary large distances in spite of the screening of the physical field.
This genuine quantum effect is purely topological and non-local in nature and can be explained in terms of the dynamics of the gauge sectors which are responsible for the long range dynamics.
Imagine now that we study the same effect but in a  time dependent background.
The corresponding topological sectors which saturate the vacuum energy will be modified due to the external background.
However, this modification can not be described in terms of any local dynamical fields, as there are no any propagating long range fields in the system since the physical electric field is screened.
The effect is obviously non-local in nature as the Aharonov-Casher effect itself is a non-local phenomenon, and cannot be expressed in terms of the local operator $F_{\mu\nu}$, but rather is expressed in terms of the gauge invariant, but non-local operator, the holonomy $\sim \exp(iQ\oint A_{\mu}dx^{\mu})$.

 We conclude this short overview  on generation of the dynamical vacuum energy  (as a result of dynamics of the topological sectors) with comment that this type of energy behaves in all respects as a cosmological constant if anomalous coupling with other gauge fields is switched off. Indeed, 
one can use conventional thermodynamical relation
\be
\label{thermodynamics}
dF=TdS-PdV, ~~~~ P=-\frac{\partial F}{\partial V}|_S
\ee
to convince yourself that the correction $\sim {\cal{T}}^{-1}$  does not modify the  equation of state. In fact, it behaves exactly in the same way  as the cosmological constant does,  i.e.
\be
\label{P-rho}
P&=&-\frac{\partial F}{\partial V}=+ \frac{32\pi^2}{g^4} \Lqcd^4  \left(1-   \frac{c_{{\cal{T}}}}{{\cal{T}}\Lqcd} \right)\nonumber\\
\rho&=&\frac{F}{V}= - \frac{32\pi^2}{g^4} \Lqcd^4  \left(1-   \frac{c_{{\cal{T}}}}{{\cal{T}}\Lqcd} \right).   
\ee
The equation (\ref{P-rho}) implies that the corresponding equation of state assumes the  form 
\be
\label{EoS}
  w &=& \frac{\Delta P}{\Delta\rho}= -1  , ~~  {\rm a}(t)\sim \exp (H t), ~~~ H\sim  {\cal{T}}^{-1},
\ee
where $\Delta P$ and $\Delta \rho$ are defined by subtracting the constant value computed in infinitely large flat  space time, as explained above and expressed by (\ref{delta}), (\ref{delta1}). 
 
 The  regime described by  (\ref{EoS}) would be the final destination of our Universe if the interaction of the QCD gauge  configurations (saturating the vacuum energy) with massless EM photons  were always switched off.  When the coupling of the QCD vacuum fields with EM field   is switched back on, the end of de Sitter behaviour  is triggered precisely by this interaction which itself is unambiguously fixed by the triangle anomaly as we discuss in next section \ref{anomalous-coupling}. 
 
 The corresponding physics of the energy transfer from the vacuum energy given by (\ref{delta}), (\ref{delta1}) to the cosmic magnetic  energy is  very similar in all respects 
 to the  physics of the reheating epoch at the end of inflation when the vacuum energy is transferred to the light gauge SM fields   as discussed in \cite{Barvinsky:2017lfl}. The 
  technical (very challenging)  problems  which need to be resolved  to address these questions are also very similar in spirit as  we   discuss  in next section.

\section{Coupling of the   vacuum energy to   photons}\label{anomalous-coupling}
This section is separated in two parts. In the first subsection \ref{tunnelling} we explain  the formal procedure (based on the Euclidean path integral formulation) which in principle allows to  compute the desired rate and other characteristics of the energy transfer.   While  the corresponding procedure  is well defined  it is not technically feasible yet. 
Therefore,  in   subsection \ref{auxiliary} we introduce an alternative technique in terms of the auxiliary topological auxiliary fields to attack the problem.

\subsection{Formulation of the problem in terms of the tunnelling transitions}\label{tunnelling}
The   vacuum energy (\ref{delta}), (\ref{delta1}) in our framework is expressed 
 in terms of the  tunnelling transitions   which are  normally computed  in terms of the Euclidean path integral and the corresponding (Euclidean)   field configurations which describe the  interpolation   between  distinct  topologically  $|k\ra$ sectors. In conventional  QFT computations the corresponding procedure selects a specific superposition of the  $|k\ra$ states which  generates the $|\theta\ra$ state with energy $E_{\rm vac}(\theta)$. In the context of DE,   when the background assumes a non-trivial FRW geometry  (in contrast with conventional  case described by $\mathbb{R}^4$)  the corresponding computations become profoundly  more complicated, though the corresponding procedure is well defined in principle:\\
1. One should describe  the  relevant Euclidean configurations satisfying the proper boundary conditions for a nontrivial geometry (similar to calorons with nontrivial holonomy,  reviewed  in Appendix A2 in  \cite{Barvinsky:2017lfl}) represented by parameter $H\sim {\cal{T}}^{-1}$ ;\\
 2. One should compute the corresponding path integral which includes all possible positions and   orientations of the relevant gauge configurations interpolating between different topological  $|k\ra$ sectors and physically describing the tunnelling transitions between them;\\
 3. The corresponding computations for the vacuum energy $\rho$ and pressure $P$ must be done with all massless fields which couple  to QCD. In our  case  the only massless particles to be considered are the    photons as production of all massive particles is exponentially suppressed. Precisely this coupling of the QCD gauge configurations with EM field is responsible for transferring the vacuum energy to the magnetic energy;\\
 4. As the last step, one should subtract the corresponding expression (computed on $\mathbb{R}^4$) as explained in previous section  \ref{topology}. Precisely this remaining portion of the vacuum energy is interpreted as the relevant energy which enters the Friedmann equation,
 and which cannot be removed by any subtraction procedure and cannot be renormalized by any UV counter terms.
 The corresponding  formulae for $\Delta\rho, \Delta P$ will depend, in general, on properties of the manifold (parametrized by $H$), the relevant coupling constant $\alpha$ with EM field, and the environment where the magnetic field is generated. This procedure will unambiguously predict the magnetic energy  of the produced  field along with  its basic features (such as the correlation length, helical features etc). 

 While these steps are well defined in principle, it is not   feasible to  perform  the corresponding  computations because even the
 first step in this direction,   a finding the relevant Euclidean configurations satisfying the proper boundary conditions for a nontrivial geometry, is yet unknown.  Nevertheless, this procedure, in principle, shows that  the de Sitter behaviour (\ref{EoS})  in this framework emerges without any local   field $\Phi(x)$ as explained  in previous section \ref{topology} because the  physics     leading to (\ref{EoS})   is not associated with  any scalar fields, but related to the tunnelling events. This procedure, in principle, also  shows how the vacuum   may transfer its energy  to the magnetic field  in a time dependent background. 
 
 In many respects this energy transfer is very similar to the so-called Dynamical Casimir Effect (DCE) when the photons are  radiated  from the vacuum in a time dependent background. The difference with conventional  DCE is that the photons are emitted in our case not from conventional  virtual fluctuating particles which always present in the system. The key difference with DCE is that  the photons in our system are emitted from vacuum configurations which describe the tunnelling processes between different topological sectors $|k\ra$. 
 
 This difference (in comparison with DCE)  in   nature of emission explicitly displays  a hard challenging technical problem in computation  of the corresponding  emission  rate. Indeed, our topological configurations  interpolating between  different topological sectors $|k\ra$ are formulated in terms of the {\it Euclidean} path integral, while the emission of real particles on mass shell   represents an inherent {\it Minkowski} process. At present time the conventional technical tools 
 developed for   Euclidean versus  Minkowski descriptions are very different and designed for different purposes and different problems. For example, conventional lattice QCD MC simulations are not designed to compute physical processes such as on shell scattering amplitudes, but perfectly adapted to compute the Euclidean correlation functions such as topological susceptibility which assumes a nonzero value exclusively due to the tunnelling events  between different topological sectors. 
 
 \subsection{Formulation of the problem in terms of the auxiliary topological fields}\label{auxiliary}

Fortunately, the key ingredients  which are relevant for our future studies can be understood  in alternative way, in terms of the  auxiliary topological non-propagating  fields  $b(x, H)$ which effectively describes the relevant infrared physics (IR) representing the key elements of the steps 1-4 highlighted   in section \ref{tunnelling}. Parameter $H$ here represents 
the deviation of the manifold under consideration (for example $1/{\cal{T}}$)  from trivial $\mathbb{R}^4$.

The basic idea is to construct the effective Lagrangian for the  auxiliary topological   field  $b(x, H)$ using the Euclidean conventional formulation. As the next step one can utilize  the standard  formulae  to rewrite  the corresponding action in Minkowski spacetime. Finally, one can study the emission of real particles and generation of real magnetic field  using the obtained effective Lagrangian written in Minkowski space. This procedure  effectively resolves  the fundamental technical problem formulated at the end of section  \ref{tunnelling} and originated  from  the   differences in descriptions  in Euclidean versus  Minkowski spacetimes.

The    formal technique we are about to overview is widely used in particle physics and condensed matter (CM) communities. We refer  to Appendix  B in ref.\cite{Barvinsky:2017lfl} for the highlights  of the main ideas and  results  of this approach  within    context of the present work.   In particular, this approach is extremely useful in description of the topologically ordered phases when the IR physics is formulated in terms of the  topological Chern-Simons (CS) like Lagrangian. One  should emphasize that the corresponding
  physics, such as calculation  of the braiding phases between quasiparticles, computation of the degeneracy etc,  can be computed (and in fact originally had been computed)  without Chern-Simons Lagrangian and without auxiliary fields. Nevertheless, the discussions of the IR physics in terms of CS like effective action  is proven to be very useful, beautiful and beneficial. In our case, it is not simply a matter of convenience, but in fact matter of necessity because 
  we cannot proceed with explicit computations along the  lines 1-4 as explained in section \ref{tunnelling}.

  In context of the present work the  auxiliary topological non-propagating  fields  $b(x, H)$ is introduced in conventional way as  Lagrange multiplier in the course of inserting  the corresponding $\delta$- functional  into the path integral which effectively constraints the relevant degrees of freedom, see Appendix  B in ref.\cite{Barvinsky:2017lfl} for references and technical  details\footnote{The computations have been performed in  a simplified version of QCD, the so-called  weakly coupled ``deformed QCD'' model \cite{Yaffe:2008} which preserves all relevant features of the strongly coupled QCD such as confinement, nontrivial $\theta $ dependence, generation of the ``non-dispersive" vacuum energy, etc  \cite{Thomas:2011ee}. The corresponding results have been reproduced
in \cite{Zhitnitsky:2013hs} using the technique of the auxiliary  topological fields $b(x)$ exploited  in  the present work.   It is expected that similar description in terms of the auxiliary topological field also holds in strongly coupled QCD. In fact,  the  the Veneziano ghost postulated in refs. \cite{ven, vendiv} can be identified with   the auxiliary 
topological fields  \cite{Zhitnitsky:2013hs}.}. The only information we need in  what follows is that 
   auxiliary field $b(x, H)$  should be thought as the source of the topological
  fluctuations, similar to the {\it axion} field, because it enters the effective Lagrangian  precisely in the same way as the   $\theta$ parameter enters the fundamental lagrangian. This   claim is explained in Appendix  B in ref.\cite{Barvinsky:2017lfl} and is based on analysis of the exact anomalous Ward Identities. In many respects  the coupling of the  $b(x, H)$ field to the gauge fields is unambiguously  determined, similar to unique coupling of the $\eta'$ field to the gluons,  photons and gauge bosons in QCD. Because we know exactly how the $\theta$ parameters couples to $E\&M$ fields we can reconstruct exactly the coupling of the auxiliary topological $b(x, H)$ field with $F_{\mu\nu}$ fields.

As a consequence of this fundamental feature  the topological auxiliary   $b(x, H)$ field is in fact an angular topological variable and it   has the same $2\pi$ periodic properties as the original $\theta$ parameter\footnote{\label{axion}As it is known the $\theta$  parameter can be promoted to the dynamical axion field $\theta(x)$ by adding the canonical kinetic term
$[\partial_{\mu}\theta(x)]^2$ to the effective Lagrangian. The difference of  the $b(x, H)$ field  with the dynamical axion $\theta(x)$ field is that the auxiliary topological field $b(x, H)$ does not have a canonical axion kinetic term.}.
 In other words, the desired coupling of $b(x, H)$  field with $F_{\mu\nu}$   photons is 
\be	\label{coup}
  {\cal L}_{b\gamma\gamma} (x)= \frac{\alpha}{4\pi} N \frac{\sum_i Q_i^2}{N_f} \left[ \theta+ b(x, H)\right] \cdot F_{\mu\nu} \tilde F^{\mu\nu} (x) \, ,
\ee
where $\alpha$ is the fine-structure constant, $Q_i$ are the electric charges of $N_f$ light quarks and    $N=3$ is the number of colours of the strongly coupled QCD and everything is written already in Minkowski metric.  As we already mentioned,
the coupling (\ref{coup}) is unambiguously fixed because the auxiliary $b(x, H)$ field always accompanies the  $\theta$ parameter in the specific combination $\left[\theta+b(x, H)\right]$ and describes the anomalous interaction of the topological auxiliary $b(x, H)$ field with $E\&M$ photons. 

The next question we want to address is as follows: 
  what are the typical fluctuation scales of the auxiliary quantum   $b(x,H)$ field? The answer is quite obvious: the typical fluctuations are of order $\Lqcd$ as  the $b(x,H)$ effectively describes the tunnelling events, and in particular,  saturates 
  the topological susceptibility (which can be explicitly computed in  weakly coupled ``deformed QCD" as studied in \cite{Zhitnitsky:2013hs} where all computations are under complete theoretical control.).   

 What happens when the same system is defined on a nontrivial manifold characterized by some dimensional parameters such as $H\sim {\cal{T}}^{-1}\ll \Lqcd$  ? In this case  the field  $b(x, H)$ will continue to fluctuate with typical frequencies $\Lqcd$. However, the relevant correlation functions
  should demonstrate
the emergence of the linear corrections with respect to these small parameters  $\sim {\cal{T}}^{-1}$. In particular,  the  topological susceptibility (expressed as the second derivative of the vacuum energy with respect to $\theta$)   should be of  order  $\Lqcd^4$   with corrections of order $(\Lqcd{\cal{T}})^{-1}$  as expressions (\ref{delta}),  (\ref{E_vac})  suggest.

It is useful to treat  $\dot{b}(x,H)$ as the axial chemical potential\footnote{\label{theta}There is a close analogy with  heavy ion physics  when  a large domain with induced  $\theta_{\rm ind}\neq 0$  can be formed resulting in generation of the  axial chemical potential  $\mu_5=\dot{\theta}_{\rm ind}$ in this  $\theta_{\rm ind}=\mu_5 t$ domain.   This term may produce a number of  interesting $P$ odd phenomena, see  \cite{Kharzeev:2009fn} for review and references.}, i.e.
\be
\label{mu_5}
 \mu_5\equiv \la \dot{b}(x,H)\ra , 
\ee
which can be easily understood by  performing the the  $U(1)_A$ chiral time-dependent transformation in the path integral to rotate away the coupling (\ref{coup}).
The corresponding interaction reappears  in the form of a singlet non-vanishing axial chemical potential $\mu_5$ for light $N_f$ flavours as stated in (\ref{mu_5}). 

Few comments are in order. In formula (\ref{mu_5})  we use notation for the expectation  value $\la \dot{b}(x,H)\ra$ to emphasize that 
 we treat $  b(x,H) $ entering (\ref{coup}) as the external parameter  ignoring a complicated  quantum dynamics of the $b(x, H) $ field itself (which would require to proceed with steps 1-4 as formulated in Sect.\ref{tunnelling}). In what follows we also   neglect  the  back-reaction of $F_{\mu\nu}$ field on $  b(x,H) $. In other words, we    approximate the dynamics of the $  b(x,H) $ by taking its 
  expectation value $ \la \dot{b}(x,H)\ra$, and treat it as an (almost constant) external thermodynamical parameter of the system.  One should emphasize that $ \mu_5$ is not a genuine thermodynamical parameter. Furthermore, $ \mu_5$ does not satisfy any classical equation of motion as there is no a canonical kinetic term in the Lagrangian for $  b(x,H) $ field itself. 
  Instead, the $b(x, H)$ field was introduced  as  Lagrange multiplier to account for complicated dynamics of the tunnelling events.

 Our next comment is related  to estimation of   the expectation value $\la \dot{b}(x,H)\ra$.  As we discussed in the previous section \ref{topology} the dimensional parameters entering our framework must be computed by  subtracting  the corresponding expectation values   computed on $\mathbb{R}^4$.  
  This procedure unambiguously implies\footnote{For this specific  case  $ \la \dot{b}(x,H=0)\ra=0$.
 Therefore, the subtraction in this case is  a triviality.}  that $\la \dot{b}(x,H)\ra\sim H$ as it must vanish at $H=0$ and it must be linear in $H$ as discussed in Sect. \ref{topology}.

Therefore, 
our problem is now   reduced to the study of the magnetic field generation determined by coupling (\ref{coup})
with a source which can be parametrized as follows 
\be
\label{coup1}
 \mu_5\equiv \la \dot{b}(x,H)\ra =c_1 H,
\ee
  where numerical coefficient $c_1\sim 1$ is order of one, similar to $c_{\cal{T}}$ from eq. (\ref{E_vac}) and it can be, in principle, computed from the first principles by
  following the steps 1-4 as highlighted in Sect. \ref{tunnelling}. 
  
  One should also remark here that many other terms may enter the right hand side (RHS)  in eq. (\ref{coup1}), depending on geometry. For example, if one considers another  geometry with extra $\mathbb{S}_z^1$ along $z$ direction one could expect the linear corrections  proportional to $\sim c_z{\cal{T}}^{-1}_z$ similar 
  to ${\cal{T}}^{-1}\sim H$ entering (\ref{coup1}). One should also expect the curvature contribution $c_RR\sim H_{\rm obs}^2$   representing  the conventional quadratic correction, see footnote \ref{H} with a comment. 
  In other words, any deviation from  $\mathbb{R}^4$, in general, contributes to RHS in eq. (\ref{coup1}).
  However, to simplify our analysis in what follows we limit ourself  with a single parameter ${\cal{T}}^{-1}\sim H$ characterizing the deviation   
  of the geometry with nontrivial holonomy\footnote{As we mentioned in Section \ref{topology} for a specific geometry studied in  \cite{Barvinsky:2017lfl} the  parameters $H$ and $\cal{T}$ are related: $H\simeq \pi/{\cal{T}}$ and describe  the  de Sitter behaviour for constant $H$.} from the topologically trivial  $\mathbb{R}^4$. We assume the $c_1H$ is the  dominating term    in eq. (\ref{coup1}).

  \exclude{
The dimensional parameter  $H$ in eq. (\ref{coup1}) should not be confused\footnote{As we mentioned in Section \ref{topology} for a specific geometry studied in  \cite{Barvinsky:2017lfl} the  parameters $H$ and $\cal{T}$ are related: $H\simeq \pi/{\cal{T}}$ and describe  the  de Sitter behaviour for constant $H$.}  with the observed (time dependent)  Hubble constant $H_{\rm obs}(t)$.
    Instead,  $H$ should be treated 
  as a parameter of the system which behaves as a cosmological constant (\ref{EoS}) saturating the DE today (\ref{delta}). 
  Numerically, these parameters are the same order of magnitude today, i.e. $H\sim  H_{\rm obs}(t)\sim 10^{-33}$ eV.
}

\section{Generation of the magnetic field through the chiral anomaly}\label{EM-generation}
\subsection{Basic equations}\label{theory}
The  coupling of the $E\&M$ fields with auxiliary topological field (\ref{coup}) parametrized by (\ref{coup1}) 
generates an additional source term in the Maxwell equations  
\be
\label{Maxwell}
\vec{\nabla}\times\vec{B}=\sigma\vec{E}+ \frac{\alpha}{2\pi} N \frac{\sum_i Q_i^2}{N_f}\cdot \left(\mu_5 \vec{B}\right),
\ee
where $\sigma$ is the conductivity to be estimated below, and  term $\sim \la \vec{\nabla} b(x,H)\ra\times \vec{E}$  was neglected as a result of    spatial isotropy of  the tunnelling events. 
The extra induced non-dissipating current $\vec{j}\sim \vec{B}$ has been a very active area of research for many years in a number of  different fields, including heavy ion physics, see reviews \cite{Kharzeev:2009fn,Kharzeev:2015znc}, the axion searches, see reviews   \cite{vanBibber:2006rb, Asztalos:2006kz,Sikivie:2008,Raffelt:2006cw,Sikivie:2009fv,Rosenberg:2015kxa,Marsh:2015xka,Graham:2015ouw,Ringwald:2016yge},  earlier studies in condensed matter physics  \cite{Alekseev:1998ds,Frohlich:2002fg}, more recent studies  in condensed matter physics \cite{Li:2014bha} to name just a few.

There are also numerous applications of this anomalous  term $\sim \mu_5 \vec{B}$ to cosmology   related to the topic of the present work, and we want to mention just   few papers \cite{Carroll:1989vb, Joyce:1997uy,Boyarsky:2011uy,Boyarsky:2015faa}   relevant for our future discussions.
 The drastic difference with most  previous studies  is that the source (\ref{coup1}) in our case is not a dynamical field, but rather, an auxiliary field accounting for   the tunnelling transitions in a time dependent background generating the vacuum dark energy (\ref{delta}), (\ref{delta1}) as discussed in Section \ref{topology}.  Nevertheless, for our purposes we can use some technical tools from previous studies  treating $\mu_5$ as almost constant thermodynamical parameter.

 One should also add that even a constant time-independent $\mu_5\neq 0$ is capable to generate the magnetic  field in the system. Indeed, the  explicit computations in cosmological context  \cite{Carroll:1989vb} and in  heavy ion collision physics   \cite{Akamatsu:2013pjd} support this claim. Our equations (\ref{B-equation}), (\ref{B(t)}) and (\ref{gamma}) below also suggest 
 that a time -independent   $\mu_5\neq 0$   generates the magnetic field. Naively, this result  may look very suspicious.
 However, one can easily see that the constant  $\mu_5\neq 0$ can be treated as  the time-dependent phase $\theta\sim \mu_5 t$, see also footnote \ref{theta}.    
 %It is known that such transformations are not in general legitimate transformations as they  violate the boundary or/and initial conditions. However, 
 This argument explicitly  shows that 
 the time dependence is, in fact, present in the system through the observable phase $\theta(t)$. Therefore,  the generation of the magnetic field for time-independent $\mu_5$ should not   surprise the readers.

 With these comments in mind we consider the following  simple ansatz for magnetic field \cite{Boyarsky:2011uy,Boyarsky:2015faa}
 \be
 \label{B} 
 \vec{B}=B(t) \big[\sin (kz), ~ \cos (kz), ~0\big],
 \ee 
while the Bianchi identity $\vec{\nabla}\times\vec{E}=-\frac{\d\vec{B}}{\d t}$ implies that the corresponding electric field assumes the form 
 \be
 \label{E} 
 \vec{E}=-\frac{1}{k}\dot{B}(t) \big[\sin (kz), ~ \cos (kz), ~0\big]=-\frac{1}{k}  \vec{\dot{B}}.
 \ee 
 The configuration (\ref{B}) is a special case of the force-free field  which satisfies
 \be
 \label{force-free}
 \vec{\nabla}\times\vec{B}=k\vec{B}, 
  \ee
see \cite{Boyarsky:2011uy,Boyarsky:2015faa} for references and   details generalizing the ansatz (\ref{B}). Substituting (\ref{B}), (\ref{E}) and (\ref{force-free})
into (\ref{Maxwell}) we arrive to the following equation for $B(t)$:
\be
\label{B-equation}
kB(t)=-\frac{\sigma}{k}\dot{B}(t)+\frac{\alpha}{\pi}\bar{c} HB(t), ~~~ \bar{c}\equiv  c_1 \frac{N\sum_i Q_i^2}{2N_f}, ~~~
\ee 
where  we introduced  $\bar{c}$ replacing the previously defined   numerical coefficient  $c_1$  as given by (\ref{coup1}).  

We are looking for a solution in the form
\be
\label{B(t)}
B(t)=B_0\exp (\gamma t),
\ee
which returns the following formula   for the exponent $\gamma$
\be
\label{gamma}
\gamma=\frac{k}{\sigma}\left[ \frac{\alpha}{\pi}\bar{c} H-k\right].
\ee
The exponential growth of the magnetic field occurs for very long waves,
\be
\label{k}
\gamma >0 ~~~  \Rightarrow  ~~~ k<  \frac{\alpha}{\pi}\bar{c} H.
\ee
The instability with respect to generation of the magnetic field $B(t)\sim \exp (\gamma t)$ due to the coupling (\ref{coup})
is well known phenomenon and was discussed previously in the literature, including the cosmological applications  \cite{Carroll:1989vb, Joyce:1997uy,Boyarsky:2011uy,Boyarsky:2015faa} and  heavy ion collisions  \cite{Akamatsu:2013pjd}. In context of the inflationary scenario the same type of coupling could be  responsible for  the reheating epoch as discussed in \cite{Barvinsky:2017lfl}. 

In context of the present work the equations (\ref{B(t)}), (\ref{k}) unambiguously imply  that the magnetic field will be generated on enormous scales of the entire  visible Universe as a result of anomalous coupling of the DE with the Maxwell field (\ref{coup}).  The generation of the magnetic field obviously implies that there will be the energy transfer from the vacuum to the magnetic  field  as a result of evolution of the Universe. 

One should emphasize that a sample  configuration 
(\ref{B}), (\ref{E}) considered above is oversimplified   example. We, of course, do not expect the magnetic field to be uniformed  running along $z$ direction through the entire Universe. Instead, we expect the field to be twisted as it is highly helical (which is normally associated with linking and twisting of the magnetic fluxes). 
Indeed, the  magnetic helicity is defined as  
\be
\label{helicity}
{\cal{H}}\equiv \int \vec{A}\cdot \vec{{{B}}} d^3x.
\ee
The time evolution of the magnetic  helicity is determined precisely by $\vec{E}\cdot \vec{{{B}}}$ entering eq.(\ref{coup}), i.e.
 \be
 \label{time}
 \frac{d{\cal{H}}}{dt}=-2\int \vec{E}\cdot \vec{{{B}}} ~d^3x.
  \ee
  For our   configuration 
(\ref{B}), (\ref{E}) considered above the   magnetic helicity  per   unit volume ${\cal{H}}/V$ is directly related to the  magnetic energy density, i.e.
\be
\label{time1}
  \frac{{\cal{H}} (t)}{V}\approx \frac{B^2(t)}{k}.
\ee
Furthermore, the time evolution of both observables is also the same as eq. (\ref{time1}) states. 
  
 One should comment here  that the magnetic field  with  enormous correlation length  is  known to be present in our Universe, see original paper \cite{neronov} and review \cite{durrer}. The mechanism suggested in the present  work automatically generates the fields with such large correlation lengths.
On other hand, it is very hard, if  at all possible, to generate such enormous correlation length  within    conventional approaches, see 
  \cite{durrer} for review. 
  
The generation  of the magnetic  field from $\mu_5$ is   not very new idea,  and was previously discussed in the literature for different systems. Furthermore, it has been known  for sometime  that the generation   of the helical magnetic field is normally  accompanied by decreasing of $\mu_5$ which is the source of the produced   field. In particular, such behaviour  is shown to occur in heavy ion systems  \cite{Akamatsu:2013pjd} and also in the  systems relevant for cosmology \cite{Boyarsky:2011uy,Boyarsky:2015faa}.

What is the efficiency of this energy transfer from the DE to the magnetic energy in our case? What is the typical time scale
for this  energy transfer? What is the intensity of the magnetic field generated by this mechanism? We have to estimate $\sigma$ and other related parameters in order to address   these and many other related questions, which is the topic  of the    next subsection.

\subsection{Numerical estimates}\label{numerics}

This subsection is much more speculative in comparison with our previous discussions in subsection \ref{theory}
which is entirely based on the Maxwell equations in the presence of additional axion  term. Nevertheless, we want to proceed
with our speculations here to argue that all conventional cosmological assumptions about the environment leads to the  estimates for the magnetic field which is perfectly consistent with presently available observations.  Future studies as discussed in  \cite{durrer}  are capable to  discover these long ranged fields. 

 We start with electric conductivity $\sigma$ entering the expression  (\ref{gamma}) for $\gamma$. It is normally estimated as follows
\be
\label{sigma}
\sigma=\frac{4\pi n_e  \alpha \tau}{m_e},
\ee
 where $\tau$ is the time scale  when a free  electron is loosing its coherence. This time scale for low density environment is normally   estimated as a result of interaction of electrons with CMB photons through the Thomson scattering
 \be
 \label{tau}
 \tau^{-1}=n_{\gamma} \sigma_T, ~~~~ \sigma_T=\frac{8\pi\alpha^2}{3m_e^2},
 \ee
 where in conventional circumstances $n_{\gamma}=n_{\rm CMB}\sim T^3$, see e.g. \cite{grasso}.  However, as we estimate  below in our framework
  the number density of the $E\&M$ configurations characterized by $\vec{B}$ and $\vec{E}$  fields and given by (\ref{B}) and (\ref{E}) correspondingly 
  is much higher than $n_{\rm CMB}$. Precisely these long-wave lengths configurations with very low $k$ as given by (\ref{k}) will be dominating  the electron resistivity in low density environment.  
   
 The estimation for the electron density $n_e$ entering (\ref{sigma}) strongly depends on the scale under consideration. For example, if residual free electrons (after recombination $p+e\leftrightarrow H+\gamma$) dominate the physics their density is estimated as  \cite{grasso}
 \be
 \label{n_e}
 n_e \approx 2\cdot 10^{-10}  (1+z)^3 {\rm cm}^{-3}.
 \ee
  At the same time if one assumes that the IGM (intergalactic medium) is mostly ionized then $n_e$ is about the average baryon density \cite{durrer}
  \be
 \label{n_e_1}
 n_e \approx \frac{\rho_B}{m_p}\approx 2\cdot 10^{-7}  (1+z)^3 {\rm cm}^{-3},
 \ee
  where $\rho_B$ is the  baryon density. 
  
  To proceed with our task we have to estimate $n_{\gamma}$ entering (\ref{tau}).
  We define the corresponding density $n_{\gamma}$ as follows
  \be
  \label{n_gamma}
  \hbar\omega_k \cdot n_{\gamma}(t)\equiv \frac{B^2(t)}{2}.
  \ee 
  For convenience of the estimates we also introduce dimensionless suppression factor  $\xi (t) <1$ which relates the magnetic energy density in comparison with the DE density, i.e. 
  \be
  \label{xi}
  \frac{B^2(t)}{2}\simeq \xi (t)  \cdot \rho_{\rm DE}(t),
  \ee
  where  $\rho_{\rm DE}(t)$ is  the source of the magnetic energy and it is defined in our framework by eqs. (\ref{delta}), (\ref{delta1}). Our goal is to estimate $\xi(t)$, and therefore, the strength of the magnetic field $B(t)$. 
  
  To achieve this goal we estimate the ratio $k/\sigma$ entering the expression for $\gamma$ in terms of the observable parameters   as follows
  \be
  \label{ratio}
  \frac{k}{\sigma}= \frac{m_e k n_{\gamma}\sigma_T}{4\pi n_e  \alpha}   \simeq \frac{\alpha}{3}\left(\frac{B^2}{n_em_e}\right)
  \simeq  \xi\frac{2\alpha}{3}\left(\frac{\rho_{\rm DE}}{n_em_e}\right)
 \ee
  where  we used previously defined relations (\ref{sigma}), (\ref{tau}), (\ref{n_gamma}), (\ref{xi}) and for estimates we take $\omega_k\approx k$. Numerically, one has
  \be
  \label{ratio1}
    \frac{k}{\sigma}\sim 2\cdot 10^2\xi\left(\frac{2\cdot 10^{-7}{\rm cm^3}}{n_e}\right).
  \ee
  This is precisely the dimensionless parameter which enters expression (\ref{gamma}) for $\gamma$. It  measures, 
   according to (\ref{B(t)}),     the typical time scale (in the Hubble units $H^{-1}$) when magnetic field  is generated. In other words,   the energy transfer from DE to the magnetic energy becomes  highly efficient when     as  $(\gamma\bar{\tau}_{\rm form})\sim 1$.   
    
  To proceed with the estimates we need to make one more assumption which is formulated as follows. It is normally assumed (in strongly coupled systems) that in order to form a configuration characterized by a typical energy $\omega$ one needs a time scale of order $(2\pi)/\omega$, which is essentially a trivial manifestation of the uncertainty relation. In our (weakly coupled case)  there is an additional fine structure coupling constant $\alpha/(2\pi)$ entering (\ref{coup}) which suggests that the time scale $\tau_{\rm form}$ required to form the magnetic configuration with wave length $k^{-1}$  from the DE source is $(2\pi/ \alpha)^2$ much longer. In addition, the time which is available for the present Universe is $H_{\rm obs}^{-1}$ which is   much shorter than $\omega_k^{-1}$ according to (\ref{k}). This  implies that $\gamma \tau_{\rm form}$ still cannot reach a magnitude  of order  one at present time;  instead, it  assumes   only a fraction of its value  $\gamma \tau_{\rm form}\sim  (\frac{\alpha\bar{c}}{\pi})\gamma\bar{\tau}_{\rm form}$ because $\frac{\omega_k}{H_{\rm obs}}\sim \frac{\alpha\bar{c}}{\pi}$
 at present time when  $H\sim H_{\rm obs}$. Collecting all these factors together we arrive to our final estimate  
  \be
  \label{form}
  &&\gamma \tau_{\rm form}\sim \frac{k}{\sigma}\left[ \frac{\alpha}{\pi}\bar{c} H-k\right] \cdot\left(\frac{2\pi}{\omega_k}\right)\cdot   \left(\frac{2\pi}{\alpha}\right)^2\\
  &\sim& 2\cdot 10^2\xi\cdot \left(\frac{2\cdot 10^{-7}{\rm cm^3}}{n_e}\right)\cdot \frac{(2\pi)^3}{\alpha^2}\approx\left(\frac{\omega_k}{H_{\rm obs}}\right)\sim \frac{\alpha\bar{c}}{\pi},\nonumber
  \ee
  where we approximated $\omega_k\sim k$ and used the  estimate (\ref{ratio1}) for $k/\sigma$. 
  
 The numerical value for $\xi$ which follows from relation (\ref{form}) can be written as follows
\be 
\label{estimate}
\xi\sim  10^{-2}\frac{\alpha^3\bar{c}}{(2\pi)^4}\left(\frac{n_e}{2\cdot 10^{-7}{\rm cm^3}}\right)\sim 10^{-12}\bar{c},
\ee  
  which implies that the intensity of the magnetic field at present time  assumes the following value 
  \be 
\label{estimate1}
B\sim \sqrt{\xi\rho_{\rm DE}}\sim   10^{-6}\cdot (2.3\cdot 10^{-3} \text{eV})^2\sim 2.6\cdot 10^{-10}G,~~~~~
 \ee 
  where we expressed  eV$^2$ units in  terms of conventional Gauss  using the following relation: $1~G\simeq 2\cdot 10^{-2} {\rm eV^2}$.
  
  This is of course, an order of magnitude estimate. However, the important point here is not just that a relatively strong magnetic field can be generated by this mechanism.  Much more important element of the proposed  mechanism is that  the source of this field  is the vacuum dark energy  $\rho_{\rm DE}$ such that these two (naively unrelated) cosmological puzzles (the nature of the dark energy and magnetic coherent field) are intimately related  because  they are both originated from the same physics governed by  the dynamics of the QCD topological sectors   as reviewed  in Sect. \ref{topology}.

  One may wonder if the generation of the magnetic field at earlier times could produce a larger intensity field
  in comparison with estimate (\ref{estimate1}). The answer is ``no", and the reason for that is as follows.  
  The wave length $ k$ is determined by eq. (\ref{k}) where  parameter $H$ is defined in (\ref{delta1}).
 At earlier times  the right hand side of equation (\ref{form}) will have an additional suppression factor $H/H_{\rm obs}\ll 1$ as the time formation in physical units is getting shorter for the same frequency $\omega_k$. This implies that parameter $\xi$, and therefore, intensity of the generated magnetic field, will receive  an additional suppression.    This argument implies that the strongest  field is generated the last.  It  should be contrasted with conventional mechanisms which could produce  very strong field at the moment of formation  but become very weak    due to the Hubble  expansion.

 \section{Conclusion and future directions}\label{conclusion}
 The main  claim of this work is that the tunnelling transitions in QCD in expanding Universe will generate the coupling (\ref{coup}) due to the chiral anomaly. This interaction unambiguously implies that the Maxwell equations will be modified
 according to eq. (\ref{Maxwell}).  This additional non-dissipating current $\vec{j}\sim \vec{B}$ in the Maxwell system implies that there will be energy transfer from the vacuum DE  to the magnetic energy. The correlation length of the  produced magnetic field   is determined by    DE  correlation   length as $B^2\sim \rho_{\rm DE}$ in this framework.   The intensity of the field generated by this mechanism is estimated on the level of $B\sim 10^{-10}$G according to  (\ref{estimate1}). 
 
Can we test some of these ideas  in  tabletop experiments, at least, as a matter of  principle?  We want to argue that the ultimate  answer is ``yes". Therefore, we claim  that we are dealing with a real physics  phenomenon,  rather  than 
with  joggling  of formal equations (such as   insertions of the Lagrange multipliers, introduction  of the auxiliary fields, subtractions of the UV counter terms and other formal elements which may look very suspicious for some readers).

The basic idea for a  tabletop experiment  goes as follows. The fundamentally new type of energy discussed in Section \ref{topology}  can be, in principle, studied  by measuring some specific corrections to the Casimir vacuum energy  in the Maxwell theory as suggested in   \cite{Cao:2013na,Zhitnitsky:2015fpa,Cao:2015uza,Yao:2016bps,Cao:2017ocv}.
This fundamentally new contribution to the Casimir pressure emerges as a result of tunnelling processes, rather than due to the conventional fluctuations of the propagating photons with two physical transverse polarizations. Therefore, it was coined as the Topological Casimir Effect (TCE).
The extra energy computed in \cite{Cao:2013na,Zhitnitsky:2015fpa,Cao:2015uza,Yao:2016bps,Cao:2017ocv} is the direct analog of the QCD non-dispersive  vacuum  energy   (\ref{delta}), (\ref{E_vac})    which is the key player of the present work as it explicitly enters the EoS (\ref{EoS}).
In fact, an extra contribution to the Casimir pressure emerges in this system as a result of nontrivial holonomy    for the Maxwell field. The nontrivial holonomy in $E\&M$ system  is  enforced by  the nontrivial boundary conditions imposed
 in refs \cite{Cao:2013na,Zhitnitsky:2015fpa,Cao:2015uza,Yao:2016bps,Cao:2017ocv}, and related to the  nontrivial mapping $\pi_1[U(1)]=\mathbb{Z}$
 relevant for  the Maxwell abelian gauge theory\footnote{A similar  new type of energy can be, in principle, also studied in superfluid He-II system which also shows a number of striking similarities
 with non-abelian QCD as argued in \cite{Zhitnitsky:2016hdz}. For the  superfluid He-II system the crucial role plays the vortices which are classified by $\pi_1[U(1)]=\mathbb{Z}$ similar to the   abelian quantum fluxes studied in  the Maxwell system.}. 
 
 Furthermore,  the emission of real physical  photons  from the Euclidean 
 vacuum configurations   describing the tunnelling events in
 the abelian Maxwell system (representing the  direct analog of the non-abelian system discussed in   Section \ref{tunnelling}) 
 can also be studied in the Maxwell theory as argued in \cite{Yao:2016bps}.

 In fact, the same  obstacle  (related to the  formulation of the tunnelling transitions  in terms of the  Euclidean path integral, while the emission of real particles on mass shell   represents   inherent  Minkowski processes) can be also resolved   by introducing the auxiliary topological fields in Maxwell system, similar to the  discussions in Section \ref{auxiliary}, see 
   \cite{Yao:2016bps} for the details.
 
  To recapitulate the main point:  the long range magnetic field with intensity of order $B\sim 10^{-10}$G can be generated as a result of variation  of the  QCD tunnelling transition rate  in the time dependent  background related to the   Universe expansion.  The  two, naively distinct phenomena, are in fact closely related as the DE is the source for the magnetic energy in this framework,  $B^2\sim \rho_{\rm DE}$. This novel effect  can be, in principle, tested in table-top experiment, and in many respects is similar to the Dynamical Casimir Effect.
  What is more important is that such fields (correlated on the enormous scale of the visible Universe) can be studied by future UHECR telescopes, see Fig. 14 in ref. \cite{durrer}. We finish this work on this optimistic note.

 \section*{Acknowledgements}
  This research was supported in part by the Natural Sciences and Engineering Research Council of Canada.


\begin{thebibliography}{99}
%\cite{Barvinsky:2017lfl}
\bibitem{Barvinsky:2017lfl} 
  A.~O.~Barvinsky and A.~R.~Zhitnitsky,
  %``Inflation and gauge field holonomy,''
  Phys.\ Rev.\ D {\bf 98}, no. 4, 045008 (2018)
  %doi:10.1103/PhysRevD.98.045008
  [arXiv:1709.09671 [hep-th]].
  %%CITATION = doi:10.1103/PhysRevD.98.045008;%%


  %\cite{Barvinsky:2006uh}
\bibitem{Barvinsky:2006uh}
  A.~O.~Barvinsky and A.~Y.~Kamenshchik,
  %``Cosmological landscape from nothing: Some like it hot,''
  JCAP {\bf 0609}, 014 (2006)
  %doi:10.1088/1475-7516/2006/09/014
  [hep-th/0605132].
  %%CITATION = doi:10.1088/1475-7516/2006/09/014;%%
  %49 citations counted in INSPIRE as of 05 Mar 2017

  %\cite{Barvinsky:2006tu}
\bibitem{Barvinsky:2006tu}
  A.~O.~Barvinsky and A.~Y.~Kamenshchik,
  %``Thermodynamics via Creation from Nothing: Limiting the Cosmological Constant Landscape,''
  Phys.\ Rev.\ D {\bf 74}, 121502 (2006)
  %doi:10.1103/PhysRevD.74.121502
  [hep-th/0611206].
  %%CITATION = doi:10.1103/PhysRevD.74.121502;%%
  %34 citations counted in INSPIRE as of 05 Mar 2017


\bibitem{why}
A. O. Barvinsky,
%``Why there is something rather than nothing (out of everything)?,''
Phys. Rev. Lett. {\bf 99} (2007) 071301 [arXIv:0704.0083[hep-th]].

 
%\cite{Zhitnitsky:2013pna}
\bibitem{Zhitnitsky:2013pna}
  A.~R.~Zhitnitsky,
  %``Inflaton as an auxiliary topological field in a QCD-like system,''
  Phys.\ Rev.\ D {\bf 89}, no. 6, 063529 (2014)
  [arXiv:1310.2258 [hep-th]].
  %%CITATION = ARXIV:1310.2258;%%

  %\cite{Zhitnitsky:2014aja}
\bibitem{Zhitnitsky:2014aja}
  A.~R.~Zhitnitsky,
  %``Cosmological perturbations in \bar{QCD}- inflation. Estimates confronting the observations, including BICEP2,''
  Phys.\ Rev.\ D {\bf 90}, no. 4, 043504 (2014)
  [arXiv:1404.5965 [hep-ph]].
  %%CITATION = ARXIV:1404.5965;%%

  %\cite{Zhitnitsky:2015dia}
\bibitem{Zhitnitsky:2015dia}
  A.~R.~Zhitnitsky,
  %``Dynamical de Sitter phase and nontrivial holonomy in strongly coupled gauge theories in an expanding universe,''
  Phys.\ Rev.\ D {\bf 92}, no. 4, 043512 (2015)
  %doi:10.1103/PhysRevD.92.043512
  [arXiv:1505.05151 [hep-ph]].
  %%CITATION = doi:10.1103/PhysRevD.92.043512;%%
  %6 citations counted in INSPIRE as of 27 Feb 2017




\bb{kronberg}
P.~P.~Kronberg,
  %``Extragalactic magnetic fields,''
  Rept.\ Prog.\ Phys.\  {\bf 57}, 325 (1994).
  %%CITATION = RPPHA,57,325;%%

\bb{grasso}
D.~Grasso and H.~R.~Rubinstein,
  %``Magnetic fields in the early universe,''
  Phys.\ Rept.\  {\bf 348}, 163 (2001)
  [arXiv:astro-ph/0009061].
  %%CITATION = PRPLC,348,163;%%


  \bibitem{durrer}
  R. Durrer and A. Neronov, Astron.\  Astrophys.\ Rev.\ {\bf 21},
62 (2013), [arXiv:1303.7121 [astro-ph.CO]].

  %\cite{Zeldovich:1967gd}
\bibitem{Zeldovich:1967gd}
  Y.~B.~Zeldovich,
  %``Cosmological Constant and Elementary Particles,''
  JETP Lett.\  {\bf 6}, 316 (1967)
  [Pisma Zh.\ Eksp.\ Teor.\ Fiz.\  {\bf 6}, 883 (1967)].
  %%CITATION = ZFPRA,6,883;%%

   %\cite{Yamamoto:2014vda}
\bibitem{Yamamoto:2014vda}
  A.~Yamamoto,
  %``Lattice QCD in curved spacetimes,''
  Phys.\ Rev.\ D {\bf 90}, no. 5, 054510 (2014)
  [arXiv:1405.6665 [hep-lat]].
  %%CITATION = ARXIV:1405.6665;%%


 %\cite{Urban:2009vy}
\bibitem{Urban:2009vy}
  F.~R.~Urban and A.~R.~Zhitnitsky,
  %``The cosmological constant from the QCD Veneziano ghost,''
  Phys.\ Lett.\ B {\bf 688}, 9 (2010)
  %doi:10.1016/j.physletb.2010.03.080
  [arXiv:0906.2162 [gr-qc]].
  %%CITATION = doi:10.1016/j.physletb.2010.03.080;%%
  %121 citations counted in INSPIRE as of 13 Sep 2017

  %\cite{Urban:2009yg}
\bibitem{Urban:2009yg}
  F.~R.~Urban and A.~R.~Zhitnitsky,
  %``The QCD nature of Dark Energy,''
  Nucl.\ Phys.\ B {\bf 835}, 135 (2010)
 % doi:10.1016/j.nuclphysb.2010.04.001
  [arXiv:0909.2684 [astro-ph.CO]].
  %%CITATION = doi:10.1016/j.nuclphysb.2010.04.001;%%
  %100 citations counted in INSPIRE as of 13 Sep 2017


 \bibitem{witten}
  E.~Witten,
  %``Current Algebra Theorems For The U(1) Goldstone Boson,''
  Nucl.\ Phys.\  B {\bf 156}, 269 (1979).
  %%CITATION = NUPHA,B156,269;%%

 \bibitem{ven}
G.~Veneziano,
  %``U(1) Without Instantons,''
  Nucl.\ Phys.\  B {\bf 159}, 213 (1979).
  %%CITATION = NUPHA,B159,213;%%

\bibitem{vendiv}
  P.~Di Vecchia and G.~Veneziano,
  %``Chiral Dynamics In The Large N Limit,''
  Nucl.\ Phys.\  B {\bf 171}, 253 (1980).
  %%CITATION = NUPHA,B171,253;%%



%\cite{Thomas:2012ib}
\bibitem{Thomas:2012ib} 
  E.~Thomas and A.~R.~Zhitnitsky,
  %``Casimir scaling in gauge theories with a gap. Deformed QCD as a toy model,''
  Phys.\ Rev.\ D {\bf 86}, 065029 (2012)
  doi:10.1103/PhysRevD.86.065029
  [arXiv:1203.6073 [hep-ph]].
  %%CITATION = doi:10.1103/PhysRevD.86.065029;%%
  %15 citations counted in INSPIRE as of 21 Jan 2019
  
\bibitem{Yaffe:2008}
  M.~\"{U}nsal and L.~G.~Yaffe,
  % "Center-stablized Yang-Mills theory: confinement and Large N Volume Independence,"
  Phys.\ Rev.\ D {\bf 78}, 065035 (2008).
  [arXiv:0803.0344 [hep-th]].


  %\cite{Thomas:2011ee}
\bibitem{Thomas:2011ee}
  E.~Thomas and A.~R.~Zhitnitsky,
  %``Topological Susceptibility and Contact Term in QCD. A Toy Model,''
  Phys.\ Rev.\ D {\bf 85}, 044039 (2012)
  [arXiv:1109.2608 [hep-th]].
  %%CITATION = ARXIV:1109.2608;%%
  %23 citations counted in INSPIRE as of 10 Jul 2014


%\cite{Zhitnitsky:2013hs}
\bibitem{Zhitnitsky:2013hs}
  A.~R.~Zhitnitsky,
  %``QCD as a topologically ordered system,''
  Annals Phys.\  {\bf 336}, 462 (2013)
  [arXiv:1301.7072 [hep-ph]].
  %%CITATION = ARXIV:1301.7072;%%
  %3 citations counted in INSPIRE as of 09 Aug 2013
  
  %\cite{Kharzeev:2009fn}
\bibitem{Kharzeev:2009fn} 
  D.~E.~Kharzeev,
  %``Topologically induced local P and CP violation in QCD x QED,''
  Annals Phys.\  {\bf 325}, 205 (2010)
  %doi:10.1016/j.aop.2009.11.002
  [arXiv:0911.3715 [hep-ph]].
  %%CITATION = doi:10.1016/j.aop.2009.11.002;%%
  %167 citations counted in INSPIRE as of 22 Jan 2019
  
  %\cite{Kharzeev:2015znc}
\bibitem{Kharzeev:2015znc} 
  D.~E.~Kharzeev, J.~Liao, S.~A.~Voloshin and G.~Wang,
  %``Chiral magnetic and vortical effects in high-energy nuclear collisions?A status report,''
  Prog.\ Part.\ Nucl.\ Phys.\  {\bf 88}, 1 (2016)
  %doi:10.1016/j.ppnp.2016.01.001
  [arXiv:1511.04050 [hep-ph]].
  %%CITATION = doi:10.1016/j.ppnp.2016.01.001;%%
  %256 citations counted in INSPIRE as of 22 Jan 2019
  
  \bibitem{vanBibber:2006rb}
  K.~van Bibber and L.~J.~Rosenberg,
  %``Ultrasensitive Searches For The Axion,''
  Phys.\ Today {\bf 59N8}, 30 (2006);


    %\cite{Asztalos:2006kz}
\bibitem{Asztalos:2006kz}
  S.~J.~Asztalos, L.~J.~Rosenberg, K.~van Bibber, P.~Sikivie, K.~Zioutas,
  %``Searches for astrophysical and cosmological axions,''
  Ann.\ Rev.\ Nucl.\ Part.\ Sci.\  {\bf 56}, 293-326 (2006).
  
    %\cite{Sikivie:2008}
\bibitem{Sikivie:2008}
  Pierre Sikivie,
  %``Axion Cosmology,'' 	
  Lect.\ Notes Phys. {\bf 741}, 19 (2008)
  % doi:10.1007/978-3-540-73518-2_2
  arXiv:0610440v2 [astro-ph].
  %%CITATION = arXiv:astro-ph/0610440v2;%%
 
%\cite{Raffelt:2006cw}
\bibitem{Raffelt:2006cw} 
  G.~G.~Raffelt,
  %``Astrophysical axion bounds,''
  Lect.\ Notes Phys.\  {\bf 741}, 51 (2008)
 % doi:10.1007/978-3-540-73518-2_3
  [hep-ph/0611350].
  %%CITATION = doi:10.1007/978-3-540-73518-2_3;%%
  %275 citations counted in INSPIRE as of 31 May 2016

  %\cite{Sikivie:2009fv}
\bibitem{Sikivie:2009fv} 
  P.~Sikivie,
  %``Dark matter axions,''
  Int.\ J.\ Mod.\ Phys.\ A {\bf 25}, 554 (2010)
  [arXiv:0909.0949 [hep-ph]].
  %%CITATION = ARXIV:0909.0949;%%

%\cite{Rosenberg:2015kxa}
\bibitem{Rosenberg:2015kxa} 
  L.~J.~Rosenberg,
  %``Dark-matter QCD-axion searches,''
  Proc.\ Nat.\ Acad.\ Sci.\  {\bf 112},  12278 (2015), 
 % doi:10.1073/pnas.1308788112
  %%CITATION = doi:10.1073/pnas.1308788112;%%
  
    
 %\cite{Graham:2015ouw}
\bibitem{Graham:2015ouw} 
  P.~W.~Graham, I.~G.~Irastorza, S.~K.~Lamoreaux, A.~Lindner and K.~A.~van Bibber,
  %``Experimental Searches for the Axion and Axion-Like Particles,''
  Ann.\ Rev.\ Nucl.\ Part.\ Sci.\  {\bf 65}, 485 (2015)
 % doi:10.1146/annurev-nucl-102014-022120
  [arXiv:1602.00039 [hep-ex]].
  
  %\cite{Marsh:2015xka}
	\bibitem{Marsh:2015xka} 
	D.~J.~E.~Marsh,
	%``Axion Cosmology,''
	Phys.\ Rept.\  {\bf 643}, 1 (2016)
	%doi:10.1016/j.physrep.2016.06.005
	[arXiv:1510.07633 [astro-ph.CO]].
	%%CITATION = doi:10.1016/j.physrep.2016.06.005;%%
	%134 citations counted in INSPIRE as of 03 Aug 2017
	

  %\cite{Ringwald:2016yge}
\bibitem{Ringwald:2016yge} 
  A.~Ringwald,
  %``Alternative dark matter candidates: Axions,''
  PoS NOW {\bf 2016}, 081 (2016)
  [arXiv:1612.08933 [hep-ph]].
  %%CITATION = ARXIV:1612.08933;%%
  %6 citations counted in INSPIRE as of 25 Dec 2017
   
 
 %\cite{Alekseev:1998ds}
\bibitem{Alekseev:1998ds} 
  A.~Y.~Alekseev, V.~V.~Cheianov and J.~Frohlich,
  %``Universality of transport properties in equilibrium, Goldstone theorem and chiral anomaly,''
  Phys.\ Rev.\ Lett.\  {\bf 81}, 3503 (1998)
  %doi:10.1103/PhysRevLett.81.3503
  [cond-mat/9803346].
  %%CITATION = doi:10.1103/PhysRevLett.81.3503;%%
  %127 citations counted in INSPIRE as of 22 Jan 2019


%\cite{Frohlich:2002fg}
\bibitem{Frohlich:2002fg}
  J.~Frohlich and B.~Pedrini,
  %``Axions, quantum mechanical pumping, and primeval magnetic fields,''
  cond-mat/0201236.
  %%CITATION = COND-MAT/0201236;%%
  
  %\cite{Li:2014bha}
\bibitem{Li:2014bha} 
  Q.~Li {\it et al.},
  %``Observation of the chiral magnetic effect in ZrTe5,''
  Nature Phys.\  {\bf 12}, 550 (2016)
  %doi:10.1038/nphys3648
  [arXiv:1412.6543 [cond-mat.str-el]].
  %%CITATION = doi:10.1038/nphys3648;%%
  %232 citations counted in INSPIRE as of 23 Jan 2019

%\cite{Carroll:1989vb}
\bibitem{Carroll:1989vb} 
  S.~M.~Carroll, G.~B.~Field and R.~Jackiw,
  %``Limits on a Lorentz and Parity Violating Modification of Electrodynamics,''
  Phys.\ Rev.\ D {\bf 41}, 1231 (1990).
  doi:10.1103/PhysRevD.41.1231
  %%CITATION = doi:10.1103/PhysRevD.41.1231;%%
  %922 citations counted in INSPIRE as of 23 Jan 2019
  
%\cite{Joyce:1997uy}
\bibitem{Joyce:1997uy} 
  M.~Joyce and M.~E.~Shaposhnikov,
  %``Primordial magnetic fields, right-handed electrons, and the Abelian anomaly,''
  Phys.\ Rev.\ Lett.\  {\bf 79}, 1193 (1997)
 % doi:10.1103/PhysRevLett.79.1193
  [astro-ph/9703005].
  %%CITATION = doi:10.1103/PhysRevLett.79.1193;%%
  %249 citations counted in INSPIRE as of 23 Jan 2019

%\cite{Boyarsky:2011uy}
\bibitem{Boyarsky:2011uy} 
  A.~Boyarsky, J.~Frohlich and O.~Ruchayskiy,
  %``Self-consistent evolution of magnetic fields and chiral asymmetry in the early Universe,''
  Phys.\ Rev.\ Lett.\  {\bf 108}, 031301 (2012)
  %doi:10.1103/PhysRevLett.108.031301
  [arXiv:1109.3350 [astro-ph.CO]].
  %%CITATION = doi:10.1103/PhysRevLett.108.031301;%%
  %118 citations counted in INSPIRE as of 22 Jan 2019
  
  %\cite{Boyarsky:2015faa}
\bibitem{Boyarsky:2015faa} 
  A.~Boyarsky, J.~Frohlich and O.~Ruchayskiy,
  %``Magnetohydrodynamics of Chiral Relativistic Fluids,''
  Phys.\ Rev.\ D {\bf 92}, 043004 (2015)
 % doi:10.1103/PhysRevD.92.043004
  [arXiv:1504.04854 [hep-ph]].
  %%CITATION = doi:10.1103/PhysRevD.92.043004;%%
  %43 citations counted in INSPIRE as of 22 Jan 2019
  
  %\cite{Akamatsu:2013pjd}
\bibitem{Akamatsu:2013pjd}
  Y.~Akamatsu and N.~Yamamoto,
  %``Chiral Plasma Instabilities,''
  Phys.\ Rev.\ Lett.\  {\bf 111}, 052002 (2013)
  [arXiv:1302.2125 [nucl-th]].
  %%CITATION = ARXIV:1302.2125;%%
  %2 citations counted in INSPIRE as of 18 Aug 2013
  

\bibitem{neronov}
  A.~Neronov and I.~Vovk,
  %``Evidence for strong extragalactic magnetic fields from Fermi observations
  %of TeV blazars,''
  Science {\bf 328}, 73 (2010)
  [arXiv:1006.3504 [astro-ph.HE]].
  %%CITATION = SCIEA,328,73;%%
  
     %\cite{Cao:2013na}
\bibitem{Cao:2013na}
  C.~Cao, M.~van Caspel and A.~R.~Zhitnitsky,
  %``Topological Casimir effect in Maxwell Electrodynamics on a Compact Manifold,''
  Phys.\ Rev.\ D {\bf 87}, no. 10, 105012 (2013)
  [arXiv:1301.1706 [hep-th]].


%\cite{Zhitnitsky:2015fpa}
\bibitem{Zhitnitsky:2015fpa}
  A.~R.~Zhitnitsky,
  %``Dynamical Casimir Effect in a small compact manifold for the Maxwell vacuum,''
  Phys.\ Rev.\ D {\bf 91}, no. 10, 105027 (2015)
  [arXiv:1501.07603 [hep-th]].

    %\cite{Cao:2015uza}
\bibitem{Cao:2015uza}
  C.~Cao, Y.~Yao and A.~R.~Zhitnitsky,
  %``Aharonov-Bohm phases in a quantum LC circuit,''
  Phys.\ Rev.\ D {\bf 93}, no. 6, 065049 (2016)
 % doi:10.1103/PhysRevD.93.065049
  [arXiv:1512.00470 [hep-th]].
  %%CITATION = doi:10.1103/PhysRevD.93.065049;%%
  %2 citations counted in INSPIRE as of 05 Mar 2017

  %\cite{Yao:2016bps}
\bibitem{Yao:2016bps}
  Y.~Yao and A.~R.~Zhitnitsky,
  %``Topological Casimir effect in a quantum LC circuit: real-time dynamics,''
  Phys.\ Rev.\ D {\bf 95}, no. 6, 065018 (2017)
  %doi:10.1103/PhysRevD.95.065018
  [arXiv:1605.01411 [hep-th]].
  %%CITATION = doi:10.1103/PhysRevD.95.065018;%%

  %\cite{Cao:2017ocv}
\bibitem{Cao:2017ocv}
  C.~Cao and A.~Zhitnitsky,
  %``Axion detection via Topological Casimir Effect,''
  Phys.\ Rev.\ D {\bf 96}, no. 1, 015013 (2017)
  %doi:10.1103/PhysRevD.96.015013
  [arXiv:1702.00012 [hep-ph]].
  %%CITATION = doi:10.1103/PhysRevD.96.015013;%%
  %3 citations counted in INSPIRE as of 11 Sep 2017


     %\cite{Zhitnitsky:2016hdz}
\bibitem{Zhitnitsky:2016hdz}
  A.~Zhitnitsky,
  %``Superfluid helium II as the QCD vacuum,''
  Nucl.\ Phys.\ B {\bf 916}, 510 (2017)
 % doi:10.1016/j.nuclphysb.2017.01.019
  [arXiv:1609.08619 [cond-mat.stat-mech]].
  %%CITATION = doi:10.1016/j.nuclphysb.2017.01.019;%%


  
 \end{thebibliography}
\end{document}